\documentclass{webofc}
\usepackage[varg]{txfonts}   
\usepackage{graphicx}
\usepackage{epsfig,amsmath}
\usepackage{amsfonts}
\usepackage{amsmath,epsfig}
\usepackage{graphicx}
\usepackage{dcolumn}
\usepackage{bm}
\usepackage{amssymb}
\usepackage{amsmath}%

\woctitle{New Frontiers in Physics 2014}
\begin{document}
\title{Influence of the measurement on the decay law: the bang-bang case}
%
%

\author{Francesco Giacosa \inst{1,2} \and Giuseppe Pagliara \inst{3}}

\institute{Institute of Physics, Jan Kochanowski University, 25-406 Kielce, Poland
\and Institute for Theoretical Physics, J. W. Goethe University,\\
Max-von-Laue-Str. 1, D--60438 Frankfurt am Main,Germany
\and Dip.~di Fisica e Scienze della Terra dell'Universit\`a di Ferrara and\\ 
INFN Sez.~di Ferrara, Via Saragat 1, I-44100 Ferrara, Italy}

\abstract{After reviewing the description of an unstable state in the framework of
nonrelativistic Quantum Mechanics (QM) and relativistic Quantum Field Theory
(QFT), we consider the effect of pulsed, ideal measurements repeated at
equal time intervals on the lifetime of an unstable system. In particular, we investigate the case in which the
`bare' survival probability is an exact exponential (a very good approximation
in both QM and QFT), but the measurement apparatus can detect the decay products only in
a certain energy range. We show that the Quantum Zeno Effect can occur in this framework as well.}
\maketitle

\section{Introduction}

The study of decays is important in atomic, nuclear, and particle physics.
Quite remarkably, although weak decays of nuclei (e.g. double-$\beta$ decays
with lifetime $\sim10^{21}$ y), and fast decays of hadrons (with lifetime
$\sim10^{-22}$ sec) are characterized by very different decay times, the basic
phenomenon is the same: a coupling of an initial unstable state to a continuum
of final states, which results in an irreversible quantum transition (infinite
Poincare' time).

It is now well established both in Quantum Mechanics (QM)
\cite{khalfin,ghirardi,mercouris} and in Quantum Field Theory (QFT)
\cite{zenoqft,duecan} that the survival probability $p(t)$ of an unstable
state $\left\vert S\right\rangle $ is never exactly an exponential function:
deviations at short as well as at long times take place. In the context of QM,
these deviations have been verified experimentally at short-times in Ref.
\cite{raizen1} and at long times in Ref. \cite{rothe}. In particular, at short
times the behavior $p(t)\simeq1-t^{2}/\tau_{Z}^{2}$ occurs. In turn, the
so-called Quantum Zeno Effect (QZE) takes place if repeated ideal measurements
at time intervals $\tau\lesssim\tau_{Z} $ are performed (bang-bang
measurements) \cite{dega,misra,facchiprl,kurizki,shimizu}. The non-decay
probability after $N$ measurements (i.e. at the time $t=N\tau$) is then given
by%
\begin{equation}
p(\tau)^{N}\simeq\left(  1-t^{2}/\tau_{Z}^{2}\right)  ^{N}\simeq\exp\left[
-\frac{t^{2}}{N\tau_{Z}^{2}}\right]  \rightarrow1\text{ for }N\rightarrow
\infty\text{ (i.e. }\tau\rightarrow0,\text{ }t\text{ fixed).} \label{qze}%
\end{equation}
Experimentally, the QZE was observed by inhibition of a Rabi oscillation
between atomic energy levels in Ref. \cite{itano,balzer} and, for a genuine
unstable quantum state (tunneling), in Ref. \cite{raizen2}.

Quite interestingly, as shown theoretically in Ref. \cite{schulman} and
experimentally verified in Ref. \cite{ketterle}, the QZE takes place also for
continuous measurements. In the latter case, a parameter, denoted as $\sigma$,
is introduced to model the coupling of the decay products with the detector (a
large value of $\sigma$ implies an intense coupling of the decay products with
the detector). The time evolution is identical to Eq. (\ref{qze}) upon the
identification $\tau=4/\sigma\ $ (the so-called Schulman relation). In turn,
in this limit the bang-bang and the continuous evolutions are indistinguishable.

Deviations from the exponential decay law are usually very small, and for this
reason they could be measured only in engineered systems. Indeed, both in QM
and in QFT the exponential decay is usually a very good approximation. There
is however an intriguing possibility to obtain the QZE also in the case of a
purely exponential decaying system: the use of an ``imperfect detector'', that
is a detector which for instance can measure the final state only if the
latter has an energy in a certain range (i.e. there is the possibility of an
undetected decay). In Refs. \cite{shimizu,Koshino:prl,koshinovari,pascapulsed}
this case has been studied theoretically by using models with a continuous
coupling of the decay products with an imperfect detector. It has been
recently re-investigated in Ref. \cite{gplast} in which the case of bang-bang
measurements has been also studied in detail. It has been found that the QZE
occurs also for repeated pulsed bang-bang measurements in which the collapse
of the wave function takes place; moreover, the outcomes of bang-bang and
continuous measurements are in general different (i.e., the Schulman relation
does not hold), a property which has interesting implications for what
concerns the interpretation of QM.

In these proceedings we first review the decay law both in QM\ and in\ QFT in
Sec. 2, where we concentrate on the derivation of the exponential limit. Then,
in Sec. 3 we study the effect of a series of ideal measurements which can
detect the final state only in a certain frequency (or energy) band. Finally,
we present our conclusions in Sec. 4.

\section{Description of an unstable state}

\subsection{QM: Lee Hamiltonian}

The Lee Hamiltonian describing the decay of an unstable state $\left\vert
S\right\rangle $ into a final state $\left\vert k\right\rangle $ is given by
$H=H_{0}+H_{1}$ with \cite{duecan,lee,giacosapra}:%
\begin{equation}
H_{0}=M_{0}\left\vert S\right\rangle \left\langle S\right\vert +\int_{-\infty
}^{+\infty}dk\omega(k)\left\vert k\right\rangle \left\langle k\right\vert
\text{ , }H_{1}=\int_{-\infty}^{+\infty}dk\frac{gf(k)}{\sqrt{2\pi}}\left(
\left\vert S\right\rangle \left\langle k\right\vert +\text{h.c.}\right)
\text{ .}%
\end{equation}
The state $\left\vert k\right\rangle $ describes the decay products,
enumerated by the continuous index $k$, which is assumed to vary between
$-\infty$ and $+\infty$. Usually, $\left\vert k\right\rangle $ represents a
two-particle state, emitted back-to-back. The survival probability amplitude
of the state $\left\vert S\right\rangle $ reads%
\begin{equation}
a(t)=\left\langle S\right\vert e^{-iHt}\left\vert S\right\rangle =\frac
{i}{2\pi}\int_{-\infty}^{+\infty}dE\left\langle S\right\vert \left[
E-H+i\varepsilon\right]  ^{-1}\left\vert S\right\rangle e^{-iEt}\text{ ,}%
\end{equation}
where in the r.h.s. the time evolution operator $e^{-iHt}$ has been
(re)written as its operatorial Fourier transform $\frac{i}{2\pi}\int_{-\infty
}^{+\infty}dE\left[  E-H+i\varepsilon\right]  ^{-1}e^{-iEt}$. The survival
probability is obtained by squaring the amplitude: $p(t)=\left\vert
a(t)\right\vert ^{2}.$ Using the equality $\left(  E-H+i\varepsilon\right)
^{-1}=\sum_{n=0}^{\infty}\left[  \left(  E-H_{0}+i\varepsilon\right)
^{-1}H_{1}\right]  ^{n}\left(  E-H_{0}+i\varepsilon\right)  ^{-1}$, the
propagator of $\left\vert S\right\rangle $ is defined and calculated as
\begin{equation}
G_{S}(E)=\left\langle S\right\vert \frac{1}{E-H+i\varepsilon}\left\vert
S\right\rangle =\frac{1}{E-M_{0}+g^{2}\Sigma(E)+i\varepsilon} \label{prop}%
\end{equation}
with the self-energy%
\begin{equation}
\Sigma(E)=-\int_{-\infty}^{+\infty}\frac{dk}{2\pi}\frac{f^{2}(k)}%
{E-\omega(k)+i\varepsilon}\text{ .} \label{sigmaqm}%
\end{equation}
If $E$ falls within a certain support $I$, the imaginary part of $\Sigma(E)$
has the explicit expression
\begin{equation}
\operatorname{Im}\Sigma(E)=\frac{1}{2}\frac{f^{2}(k_{E})}{\left(
\frac{d\omega}{dk}\right)  _{k=k_{E}}}\text{ }%
\end{equation}
where $k_{E}$ is defined by $\omega(k_{E})=E$. If $E$ lies outside $I,$ the
imaginary part simply vanishes. The (renormalized) nominal mass $M$ of the
state $\left\vert S\right\rangle $ is defined as the solution of the
equation:
\begin{equation}
M-M_{0}+g^{2}\operatorname{Re}\Sigma(M)=0\text{ .}%
\end{equation}
By expanding the real part of $G_{S}^{-1}(E)$ around $M$ we obtain the
Breit-Wigner approximation of the propagator (by eliminating an inessential
constant) as%
\begin{equation}
G_{S}^{\text{BW}}(E)=\frac{1}{E-M+i\Gamma_{\text{BW}}/2} \label{gsbw}%
\end{equation}
where the decay width $\Gamma_{\text{BW}}$ is given by a generalization of the
Fermi golden rule:
\begin{equation}
\Gamma_{\text{BW}}=\frac{g^{2}}{1+g^{2}\left(  \frac{\partial\operatorname{Re}%
\Sigma(E)}{\partial E}\right)  _{E=M}}\frac{f^{2}(k_{M})}{\left(
\frac{d\omega}{dk}\right)  _{k=k_{M}}}\text{ }%
\end{equation}
and where $k_{M}$ is given by $\omega(k_{M})=M.$ Then, by using the
approximation (\ref{gsbw}) the survival probability amplitude is given by
$a(t)=\left\langle S\right\vert e^{-iHt}\left\vert S\right\rangle
=e^{-i(M-i\Gamma_{\text{BW}}/2)t}$ and the survival probability takes the
usual form $p(t)=\left\vert a(t)\right\vert ^{2}=e^{-\Gamma_{\text{BW}}t}.$

For illustrative purposes we present a simple model in which two simplifying
assumptions for the functions $\omega(k)$ and $f(k)$ are made
\cite{procbregrenz,thomas,gp}: (i) $\omega(k)=k.$ This choice implies that the
energy is not bounded from below. Thus, it is obviously an -indeed a quite
general- approximation, which is also needed to derive the usual exponential
decay. (ii) $f(k)=\theta(M_{0}+\Lambda-k)\theta(k-(M_{0}-\Lambda))$. In this
way, the unstable state $\left\vert S\right\rangle $ couples in a limited
$\emph{window}$ of energy to the final states of the type $\left\vert
k\right\rangle $. Conditions (i) and (ii) define a toy model for the decay law
in the general framework of Lee Hamiltonians. The self-energy $\Sigma(E)$
reads:
\begin{equation}
\Sigma(E)=\frac{1}{2\pi}\ln\left(  \frac{E-M_{0}+\Lambda}{E-M_{0}-\Lambda
}\right)  \text{ ,}%
\end{equation}
whose real and imaginary parts are $\operatorname{Re}\Sigma(E)=\frac{1}{2\pi
}\ln\left\vert \frac{E-M_{0}+\Lambda}{E-M_{0}-\Lambda}\right\vert $ and
$\operatorname{Im}\Sigma(E)=\frac{1}{2}$ for $M_{0}-\Lambda<E<M_{0}+\Lambda$,
$0$ otherwise.

When $\Lambda$ is finite, deviations both at short and long times occur, see
the explicit numerical results in \cite{procbregrenz,gp,thomas}. However, in
the limit $\Lambda\rightarrow\infty$ the cutoff model reduces exactly to the
BW case and thus to the pure exponential decay. Namely, $\operatorname{Re}%
\Sigma(E)=0$ and $\operatorname{Im}\Sigma(E)=g^{2}/2$ for each value of $E$
and the propagator reads $G_{S}(E)=\left[  E-M+i\Gamma/2\right]  ^{-1}$ with
$M=M_{0}$ and $\Gamma=\Gamma_{\text{BW}}=g^{2}$. The time-evolution operator
applied to the unstable state $\left\vert S\right\rangle $ delivers the
following result \cite{giacosapra,ww,scully,facchispont}:
\begin{align}
e^{-iHt}\left\vert S\right\rangle  &  =e^{-i(M_{0}-i\Gamma/2)t}\left\vert
S\right\rangle +\int_{-\infty}^{+\infty}dkb(k,t)\left\vert k\right\rangle
\text{ }\nonumber\\
\text{with }b(k,t)  &  =\frac{g}{\sqrt{2\pi}}\frac{e^{-ikt}-e^{-i(M_{0}%
-i\Gamma/2)t}}{k-M_{0}+i\Gamma/2}\text{ .} \label{expev}%
\end{align}
The survival probability amplitude is $a(t)=\left\langle S\right\vert
e^{-iHt}\left\vert S\right\rangle =e^{-i(M_{0}-i\Gamma/2)t}$, from which
$p(t)=e^{-\Gamma t}$. Notice that $e^{-iHt}\left\vert k\right\rangle
=e^{-ikt}\left\vert k\right\rangle $ holds in the present case.

\subsection{QFT: relativistic Lagrangian}

In the framework of relativistic QFT one obtains a picture similar to the
QM\ case. In order to recall the main features we use the following Lagrangian
with the scalar fields $S$ and $\varphi$ \cite{zenoqft,duecan,lupo}:%
\begin{equation}
\mathcal{L}=\frac{1}{2}(\partial_{\mu}S)^{2}-\frac{1}{2}M_{0}^{2}S^{2}%
+\frac{1}{2}(\partial_{\mu}\varphi)^{2}-\frac{1}{2}m^{2}\varphi^{2}%
+gS\varphi^{2}\text{ .}%
\end{equation}
The decay process $S\rightarrow\varphi\varphi$ is analogous to the transition
$\left\vert S\right\rangle \rightarrow\left\vert k\right\rangle $ of the
previous subsection. The propagator of the particle $S$ is obtained upon Dyson
resummation as
\begin{equation}
G_{S}(p^{2})=-i\int d^{4}ye^{-ip\cdot y}\left\langle 0\left\vert T\left[
S(0)S(y)\right]  \right\vert 0\right\rangle =\frac{1}{p^{2}-M_{0}^{2}%
+(\sqrt{2}g)^{2}\Sigma(p^{2})+i\varepsilon\text{ }} \label{sprop}%
\end{equation}
where $T$ is the chronological product, $\left\vert 0\right\rangle $ the
(non-perturbative) vacuum state, and $\Sigma(p^{2})$ the self-energy .\ The
form of Eq. (\ref{sprop}) is analogous to Eq. (\ref{prop}). in the one-loop
approximation (which is usually a very good one in this context, see the
recent Ref. \cite{schneitzer}.) $\Sigma(p^{2})$ reads \cite{duecan,lupo}:
\begin{equation}
\Sigma(x^{2})=\int\frac{d^{3}q}{(2\pi)^{3}}\frac{\tilde{\phi}^{2}(\mathbf{q}%
)}{\sqrt{\mathbf{q}^{2}+m^{2}}\left(  4(\mathbf{q}^{2}+m^{2})-x^{2}%
-i\varepsilon\right)  }\text{ .}%
\end{equation}
which has a form very similar to Eq. (\ref{sigmaqm}) (for a more detailed
comparison of QM and QFT, see Ref. \cite{duecan}). The choice of the
vertex-function $\tilde{\phi}(\mathbf{q})$ depends on the model and on the
physics that one wants to study \cite{lupo,altri}. For illustrative purposes,
the cutoff choice $\tilde{\phi}(\mathbf{q})=\theta(\mathbf{q}^{2}-\Lambda
^{2})$ is useful and represents the analogous expression of the `cutoff model'
of the QM case.

The renormalized mass $M$ of the particle $S$ is defined as the zero of the
real part of $G_{S}(p^{2})^{-1}:$ $M^{2}-M_{0}^{2}+\operatorname{Re}\Pi
(M^{2})=0.$ Usually $\operatorname{Re}\Pi(M^{2})>0$, thus $M$ is smaller than
$M_{0}.$ The tree-level decay width of the process $S\rightarrow\varphi
\varphi$ reads:%
\begin{equation}
\Gamma=\Gamma^{\text{tl}}(x=M)\text{ , }\Gamma^{\text{tl}}(x)=\frac
{\sqrt{\frac{x^{2}}{4}-m^{2}}}{8\pi x^{2}}\left(  \sqrt{2}g\right)  ^{2}%
\theta(x-2m)\text{ .}%
\end{equation}
In order to obtain the BW expression of the propagator two steps are
necessary.\ First, one simplify $G_{S}$ via the so-called relativistic BW
approximation as

\[
G_{S}\simeq\frac{N_{\text{BWrel}}}{x^{2}-M^{2}+iM\Gamma}\text{ }%
\]
where $N_{\text{BWrel}}$ is a renormalization constant. Then, one obtains the
non-relativistic case in the following way:%
\begin{equation}
G_{S}\simeq\frac{N_{\text{BWrel}}}{\left(  x-M\right)  (2M)+iM\Gamma}%
\simeq\frac{N_{\text{BW}}}{x-M+i\Gamma/2}\text{ .}%
\end{equation}
Thus, we obtain from the relativistic framework also the nonrelativistic BW
propagator as a special limit. It is indeed remarkable that this limit works
so well also for strong decaying particles, as for instance for the $\rho$ meson.

\section{Pulsed measurements with an `imperfect' detector}

Let us now consider the case in which the intrinsic decay is exactly an
exponential (i.e., it is assumed that the small deviations from the
exponential decay are completely negligible for our purposes.) We aim to show
which is the effect of a series of imperfect measurements at time intervals
$\tau,$ $2\tau,$ ..., $t=n\tau.$ The discussion is taken from Ref.
\cite{gplast}, to which we add some technical details of the derivation of the
main formula of that paper.

In the exponential limit Eq. (\ref{expev}) holds. Without loss of generality,
we set $M_{0}=0$. The state of the system at the instant $t>0$ is given by
$\left\vert \Psi(t)\right\rangle =e^{-iHt}\left\vert S\right\rangle $ (valid
up to the the first measurement). Then, we assume to perform a measurement at
the instant $\tau$ and that the detector $D$ can only detect the final state
$\left\vert k\right\rangle $ if $-\lambda\leq k\leq\lambda.$ This means that
the probability to \textquotedblleft hear\textquotedblright\ the click of the
detector is given by:
\begin{equation}
p_{\text{click}}(\tau)=w_{\lambda}(\tau)=\int_{-\lambda}^{\lambda}\left\vert
b(k,\tau)\right\vert ^{2}dk\text{ .}%
\end{equation}
Conversely, the no-click probability after the first measurement is
$p_{\text{no-click}}(\tau)=1-w_{\lambda}(\tau).$ (Note, for $\lambda
\rightarrow\infty$ the probability that the detector makes click for a single
measurement at the instant $\tau$ is $w_{\infty}(\tau)=\int_{-\infty}%
^{+\infty}dk\left\vert b(k,\tau)\right\vert ^{2}=1-p(\tau)=1-e^{-\Gamma t}$,
as expected.) Then, if the detection has not taken place, the state of the
system at the instant of time $\tau^{+}$ (just after the first measurement) is
given by%
\begin{equation}
\left\vert \Psi(\tau^{+})\right\rangle =N\left[  e^{-\tau\Gamma/2}\left\vert
S\right\rangle +\int_{R}dkb(k,\tau)\left\vert k\right\rangle \right]
\end{equation}
where $R=(-\infty,-\lambda)\cup(\lambda,\infty)$ and $N=1/\sqrt
{p_{\text{no-click}}(\tau)}$ is a normalization constant arising after that
the wave function has collapsed (the band $W=(-\lambda,\lambda)$ has
disappeared from the linear superposition because no click has occurred).
Further unitary time-evolution implies that for $t>\tau$ (valid up to the
second measurement):
\begin{equation}
\left\vert \Psi(t=\tau+t^{\prime})\right\rangle =N\left[  e^{-\tau\Gamma
/2}\left(  e^{-t^{\prime}\Gamma/2}\left\vert S\right\rangle +\int_{-\infty
}^{+\infty}dkb(k,t^{\prime})\left\vert k\right\rangle \right)  +\int
_{R}dkb(k,\tau)e^{-ikt^{\prime}}\left\vert k\right\rangle \right]  \text{ .}%
\end{equation}
Note, the normalization of the state for $t>\tau$ implies that%
\begin{equation}
\left\vert N\right\vert ^{2}\left[  p(\tau)p(t^{\prime})+p(\tau)w_{\lambda
}(t^{^{\prime}})+X(t^{^{\prime}})\right]  =1\text{ valid }\forall t^{^{\prime
}}\text{ }\label{normtprime}%
\end{equation}
with $X(t^{\prime})=\left\vert \int_{R}dk\left[  b(k,t^{\prime}%
)e^{-ikt^{\prime}}+e^{-\tau\Gamma/2}b(k,t^{\prime})\right]  \right\vert ^{2}$.
Now, we assume to make a second measurement at the time $t=2\tau.$ The
probability to hear a click at the time $2\tau$ is given by $p_{\text{click}%
}(2\tau)=p_{\text{no-click}}(\tau)\left\vert N\right\vert ^{2}p(\tau
)w_{\lambda}(\tau)=p(t)w_{\lambda}(\tau).$ Conversely, if no click takes place
also at $t=2\tau,$ just slightly after it we have:
\begin{equation}
\left\vert \Psi(t=2\tau^{+})\right\rangle =\tilde{N}\left[  e^{-\tau\Gamma
/2}\left(  e^{-\tau\Gamma/2}\left\vert S\right\rangle +\int_{R}dkb(k,\tau
)\left\vert k\right\rangle \right)  +\int_{R}dkb(k,\tau)e^{-ik\tau}\left\vert
k\right\rangle \right]  \text{ }%
\end{equation}
where $\tilde{N}$ is a new normalization constant after the second collapse.
By using Eq. (\ref{normtprime}) one obtains:%
\begin{equation}
\left\vert \tilde{N}\right\vert ^{2}\left[  p(\tau)^{2}+X(\tau)\right]
=1\rightarrow\left\vert \tilde{N}\right\vert ^{2}\left[  \frac{1}{\left\vert
N\right\vert ^{2}}-p(\tau)w_{\lambda}(\tau)\right]  =1\text{ ,}%
\end{equation}
from which:%
\begin{equation}
\left\vert \tilde{N}\right\vert ^{2}=\frac{1}{1-w_{\lambda}(\tau
)-p(\tau)w_{\lambda}(\tau)}\text{ .}%
\end{equation}
Then, one can repeat the procedure as often as desired. The next step is a
measurement at the time $t=3\tau,$ and so on. The outcome is that the
probability to hear a click exactly at the $n$-th measurement is given by:
\begin{equation}
p_{\text{click}}(n\tau)=p^{n-1}(\tau)w_{\lambda}(\tau)\text{ .}%
\label{recurring}%
\end{equation}
This formula is intuitive, as it expresses the probability to hear click at
the $n$-th measurement as the joint probability that the state $\left\vert
S\right\rangle $ is still present at the time $(n-1)\tau$ and that the decay
takes place exactly in the very last step. From Eq. (\ref{recurring}) we
calculate the no-click probability up to the instant $t=n\tau$ as:
\begin{align}
p_{\text{no-cklick}}(t &  =n\tau)=1-\sum_{k=1}^{n}p_{\text{click}}%
(k\tau)=1-\sum_{k=1}^{n}p^{k-1}(\tau)w_{\lambda}(\tau)\nonumber\\
&  =1-w_{\lambda}(\tau)\frac{1-p(\tau)^{n}}{1-p(\tau)}=1-w_{\lambda}%
(\tau)\frac{1-e^{-\Gamma t}}{1-e^{-\Gamma\tau}}\text{.}\label{pnc}%
\end{align}
This is indeed one of the main results of Ref. \cite{gplast}, which we have
here derived with more details.

In Fig.1, we display two numerical examples of Eq. (\ref{pnc}) (dashed lines)
corresponding to two values of $\tau$ in comparison with the exponential
survival probability (orange lines). Notice that the evolution of the system
is strongly slowed down due to the repeated measurements: after a time period
of 5 times the lifetime of the unstable system, the no-click probability is
still of the order of 90\% in the left panel and 20\% in the left panel. This
is the usual QZE which indeed takes place also in presence of exponentially
decaying systems but which are "observed'' by imperfect detectors. Another
interesting feature is that the no-click probability saturates to a finite
value at late times. In the figure we also show results for continuous
measurements (solid lines) with $\sigma$ related to $\tau$ by the Schulman
relation \cite{schulman}. Contrary to the case of perfect detectors, in the
case of imperfect ones the results for continuous and pulsed measurements
differ from each other. As discussed in Ref. \cite{gplast}, the qualitative
difference between these two types of measurements could in principle be
exploited for clarifying whether the collapse of the wave function, which has
been explicitly constructed previously for deriving equation (\ref{pnc}), is a
real physical process playing a fundamental role for the QZE.

\begin{figure}[ptb]
\vskip 0cm \begin{centering}
\epsfig{file=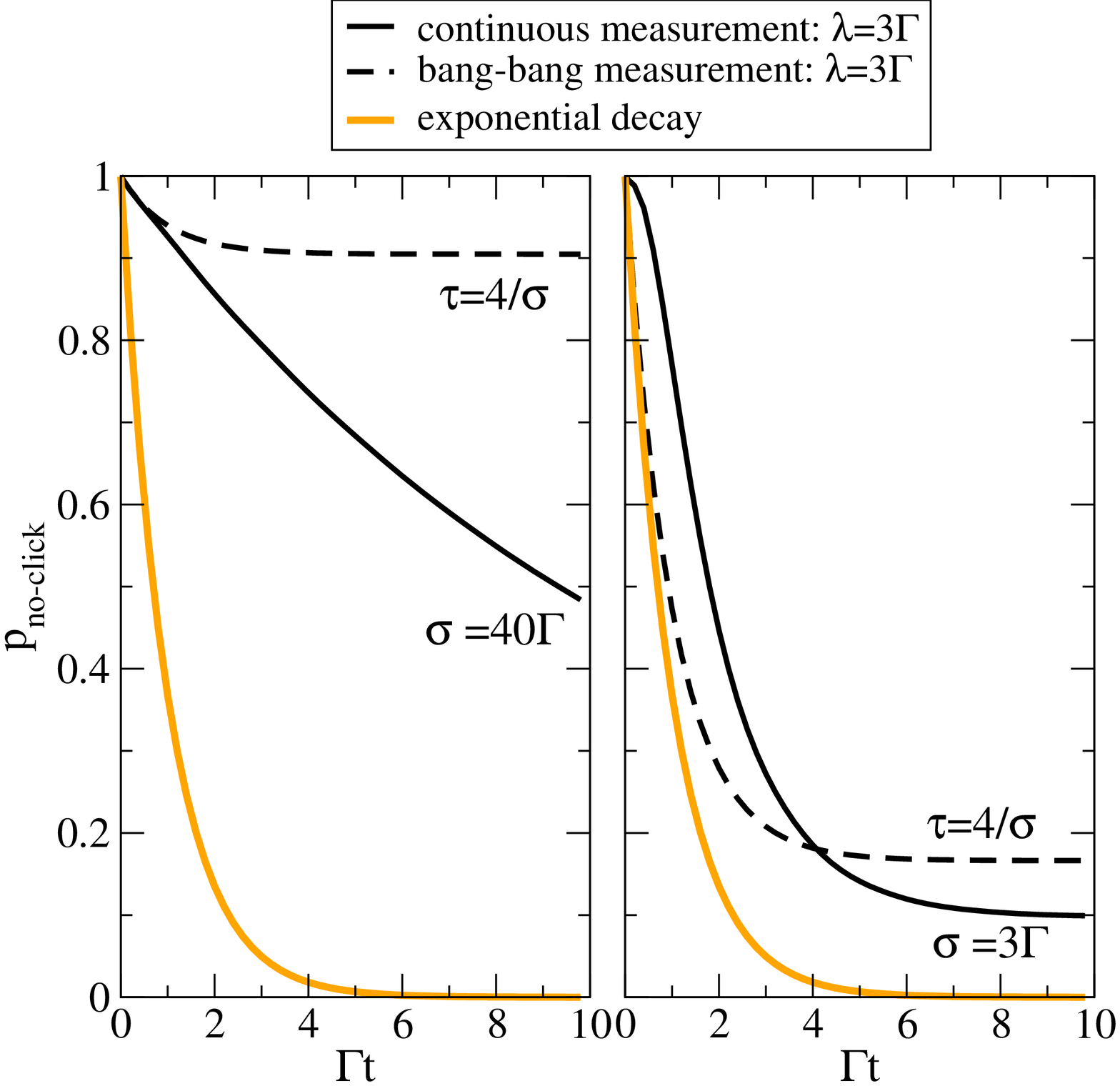,height=8.5cm,width=6cm,angle=-90}
\caption{Temporal evolution of the no-click probabilities in the case of exponential decay, continuous and pulsed measurements
for two values of the efficiency of detection $\sigma$ (from \cite{gplast}).}
\end{centering}
\end{figure}

\section{Discussions and conclusions}

In this work we have first reviewed the emergence of the Breit-Wigner limit,
and conversely of the exponential decay, in both QM and QFT. To this purpose
we have studied the propagators in QM and QFT and presented the analogies
between them. Quite interestingly, the exponential limit, although never
exact, works often very well in both cases.

As a next step we concentrated our attention to the case in which the decay
law is an exact exponential \emph{but} the detector can detect the decay
products only in a certain range of energy (it is in a sense an ``imperfect
detector'' since in some cases the decayed particles pass undetected). In this
case, even if the intrinsic decay is treated as an exact exponential function,
the quantum Zeno effect can be induced by frequent measurements of the ideal
type (collapse of the wave function after each measurement). We have shown how
the probability of not hearing a click (which is not equal to the survival
probability because the detector has not a 100\% efficiency) is derived and
some numerical example has been also provided.

In the future, one should investigate further a detector modelled by a series
of ideal but imperfect measurements by taking into account that they do not
take place at equal time intervals. Further comparison with the alternative
view of a continuous measurement is interesting, see Ref. \cite{gplast} and
references therein. This comparison has also important implications for
fundaments of QM, i.e. if the collapse of the wave function is a real physical
process as proposed in Refs. \cite{bassighirardi,bassi,grw,penrose,qmfra}, or
not, see e.g. \cite{everett,dsw}.

\end{document}